\documentclass[rnaas]{aastex62}

\begin{document}

\title{The 1D Relativistic Doppler Formula is an Incorrect Approximation in Precise Radial Velocity Work}

\correspondingauthor{Jason T.\ Wright}
\email{astrowright@gmail.com}
\author[0000-0001-6160-5888]{Jason T.\ Wright}
\affil{Department of Astronomy \& Astrophysics / 525 Davey Laboratory /
The Pennsylvania State University /
University Park, PA, 16802, USA}
\affil{Center for Exoplanets and Habitable Worlds / 525 Davey Laboratory /
The Pennsylvania State University /
University Park, PA, 16802, USA}


\section{} 

Stellar Doppler velocimetry determines a star's radial velocity $v_r$ via measurement of a redshift, $z$. At precisions below 10 m s$^{-1}$ conversion between the two quantities is complex, and care must be taken to properly account for the effects of relativity. \citet{Lindegren03} and \citet{bary} provide expositions of many of the issues involved in transforming $z$ to $v_r$ in the barycentric frame.

Often, such rigorous treatments are inappropriate because one simply needs to express a Doppler shift in units of velocity, without attributing any specific part of it to the motion of the source or observer. I write this note to emphasize that the 1D relativistic equation, despite its appearance in much public and private software for the purpose,  is incorrect in this context, and potentially the source of significant errors.

The classical Doppler shift is expressed in terms of wavelength $\lambda$ for $v_r \ll c$ as 
\begin{equation}
\label{classical}
z \equiv \frac{\Delta\lambda}{\lambda_{\rm emitted}} \approx \beta_r \equiv \frac{v_r}{c}
\end{equation}
\noindent where the $r$ subscript refers to a vector's radial component.\footnote{This is the standard in optical work. Radio astronomy, concerned more with frequency than wavelength, historically used the convention $z = \beta_r/(1-\beta_r)$. See \citet{Lindegren03}.}

The effects of relativity enter the problem in constructing the appropriate frame in which one defines $v$, defining the radial vector $\hat{r}$, and in the second- and higher-order $\beta$ terms that correct the classical formula. Earth's orbital motion in the frame of the solar system barycenter (SSB) is of order $\beta_\Earth\approx10^{-4}$, so relativistic effects on velocity measurements of astronomical sources made from Earth matter at the $c\beta_\Earth^2 \approx 3$ m s$^{-1}$ level. Appropriate relativistic corrections are thus an essential part of precise Doppler velocimetry for exoplanet detection and characterization.

In some code, these corrections are misapplied through use of the one-dimensional form of the relativistic Doppler equation:
\begin{equation}
z_{\rm 1D} = \gamma(1+\beta_r) -1 = \frac{1+\beta_r}{\sqrt{1-\beta^2}} - 1  \label{1D}
\end{equation}
Since, in 1D, $\beta=\beta_r$, this simplifies to the common form of the 1D Doppler formula, which does not distinguish between the motion of the source and that of the observer:
\begin{eqnarray}
z_{\rm 1D}&=& \sqrt{\frac{1+\beta_r}{1-\beta_r}} -1 \label{wrong}\\
& \approx & \beta_r + \frac{1}{2}\beta_r^2\label{approx}
\end{eqnarray}

This is an incorrect application of that equation because it ignores the transverse redshift, which for observers on Earth is of the same magnitude as the radial relativistic correction. The cancellation that produces Equation~\ref{wrong} from Equation~\ref{1D} is thus invalid. It also elides the complexities in defining the appropriate radial vector, and of distinguishing between the effects of the motion of the source and the observer.

These are not minor effects: in reality the Lorentz factor for Earth's motion is approximately constant, but in the 1D formula it contributes all of the variation in the second-order term, introducing errors of order 3 m s$^{-1}$. To wit, the special relativistic redshift formula for a moving star as observed from the Earth in terms of velocities and directions measured in the frame of the SSB is \citep[simplified\footnote{The full solution including perspective effects, gravitational redshift, and other quantities should be used for work below 1 m s$^{-1}$.} from equation 24 of][]{bary}:
\begin{equation}
\label{correct}
z = \frac{\gamma_*(1+\beta_{*,r})}{\gamma_\Earth(1+\beta_{\Earth,r})}-1 
 \approx (\beta_{*,r}-\beta_{\Earth,r}) + \frac{1}{2}(\beta_{*,r}-\beta_{\Earth,r})^2 + \frac{1}{2}(\beta_{*,T}^2-\beta_{\Earth,T}^2)
\end{equation}
\noindent where the $T$ subscript refers to the transverse component of $\beta$, and $\Earth$ and $*$ refer to the observer and source, respectively. I write the expression in this form to highlight its relationship to Equation~\ref{approx}, and how that equation misses the significant contribution from the transverse component of the Earth's motion.

Since in most applications the Lorentz factors are very nearly constant (contributing variation only at the $\sim 3$ cm s$^{-1}$ level) it is more useful to recast Equation~\ref{correct} as:

\begin{equation}
z = \frac{\gamma_*(1+\beta_{*,r})}{\gamma_\Earth(1+\beta_{\Earth,r})}-1 \approx (\beta_{*,r} - \beta_{\Earth,r})-\beta_{\Earth,r}(\beta_{*,r} - \beta_{\Earth,r}) +[\beta_*^2 - \beta_\Earth^2]/2
\end{equation}

\noindent where the term in brackets is roughly constant. This form makes it clear that for practical purposes, the 1D formula does not even provide the correct coefficient for variations in the second-order terms involving $\beta_{\Earth,r}$.

In code where one needs to report a redshift in the more familiar units of velocity without a rigorous calculation, it is important to clearly document which formula is being used, and I recommend simply (and explicitly) using the approximation $v_r \approx cz$. This choice is trivially inverted, does not misrepresent the degree of relativistic rigor that has been applied in translating between redshift and radial velocity, and is perhaps the most commonly followed convention in astronomy and cosmology.

Only when reporting the precise, measured changes in a star's motion, as in exoplanet detection, must one apply a full barycentric correction. In such cases one stars with the measured redshift $z$ and follows a barycentric correction procedure, such as the prescription of \citet{bary} or the IAU definition of a spectroscopic {\it barycentric radial velocity measure} \citep{Lindegren03}.\footnote{The primary differences between the two are that the \citet{bary} procedure corrects measured radial velocities for the Shapiro delay in the Solar System and the changing light travel time to the star due to its space motion; and that \citet{bary} remove two ``nuisance'' terms due to changing perspective on the star, namely the secular acceleration and the parallax-proper-motion cross term. When comparing the procedures, note the unfortunate choice of \citet{bary} to use the symbol $z_B$ to refer to the redshift induced by the observatory's motion about the SSB after \citet{Lindegren03} had already chosen that symbol to refer to the barycentric-motion-corrected redshift of the star.} 

For other purposes (for instance when writing software to precisely measure or apply redshifts, reporting the measured redshift that includes the Earth's motion, or reporting a ``barycentric correction'' value) it is best to refer only to the dimensionless redshift---not a ``radial velocity''---because it is the actual quantity being applied, measured, or reported. Expressing these redshifts as velocities introduces ambiguity in exactly which formula has been used (e.g.\ the ``optical'' classical formula, ``radio'' classical formula, 1D relativistic formula, the \citet{bary} procedure, or the IAU standard reduction).

\acknowledgments
I thank Suvrath Mahadevan for soliciting this note.


\begin{thebibliography}{}
\expandafter\ifx\csname natexlab\endcsname\relax\def\natexlab#1{#1}\fi
\providecommand{\url}[1]{\href{#1}{#1}}

\bibitem[{{Lindegren} \& {Dravins}(2003)}]{Lindegren03}
{Lindegren}, L., \& {Dravins}, D. 2003, \aap, 401, 1185

\bibitem[{{Wright} \& {Eastman}(2014)}]{bary}
{Wright}, J.~T., \& {Eastman}, J.~D. 2014, \pasp, 126, 838

\end{thebibliography}
\end{document}